\documentstyle[pre,aps]{revtex}
\begin{document}
\input{epsf.tex}
\epsfverbosetrue

\title{Instabilities of vortices in binary mixture of trapped Bose-Einstein condensates: Role of collective excitations with positive and negative energies 
}
\author{
Dmitry V. Skryabin 
\footnote{URL:
www://http.cnqo.strath.ac.uk/$\sim$dmitry}}

\address{Department of Physics and Applied
Physics, University of Strathclyde, Glasgow, G4 0NG, United Kingdom}

\date{19 January, 2000; to appear in Phys. Rev. A, January 2001}

\maketitle

\begin{abstract}
Correspondence between frequency and energy spectra and biorthogonality conditions  for the 
excitations of  Bose-Einstein condensates  described by Gross-Pitaevskii model
have been derived  selfconsistently
revealing several novel aspects  originating in nonselfadjoitness of the
Bogoliubov operator. It has been demonstrated that 
frequency resonances of the excitations  with positive  and negative energies can lead to
their mutual annihilation  and appearance of the collective modes with complex frequencies
and zero energies. Conditions for the avoided crossing of energy levels have also been
discussed. General theory has been verified  
both  numerically and analytically in the  weak
interaction limit considering an example of vortices in a binary mixture of condensates.
Growth of excitations with complex frequencies leads  to 
  spiraling of the unit and double vortices out of the
condensate center to its periphery and to  splitting of the double and higher order vortices to the unit ones.  
\end{abstract}

%\begin{multicols}{2}
%\narrowtext
%******************************************************
\section{Introduction}
%*****************************************************
Recent observations of the quantized vortices in a dilute ultracold gas of  $^{87}Rb$ atoms
\cite{matthews,madison}  are spectacular evidences of the  superfluid
properties  of  atomic gases  below  transition temperature for  formation of Bose-Einstein
condensate (BEC). First experimental results were obtained, using method suggested by
Williams and Holland \cite{Williams}, in a binary mixture of the  hyperfine states of
$^{87}Rb$ \cite{matthews} and  more recent experiments \cite{madison}  have demonstrated
vortices and vortex lattices in an optically stirred single component condensate.

Gross-Pitaevskii (GP) equation  \cite{pit} is a widely used approximation
to describe dynamics of the quantized vortices in the superfluid component of Bose gases at temperatures below critical.
The GP equation falls into  a general class of Hamiltonian nonlinear  systems. The theory
of stability of stationary solutions (equilibria) 
of such systems is well established at the moment,  see e.g.
\cite{mackay}. However, a certain gap between formal mathematical knowledge and
applications to physical examples still exists and BEC field is not an exception. One of
the controversial examples in the BEC context is complex or imaginary eigenfrequencies in the spectrum of elementary excitations.  Existence of these frequencies
was first pointed out by  Bogoliubov in his original work \cite{bog}, for more recent references see, e.g.
\cite{goldstein,phasesep,pu99,ripoll1,ripoll2,federnew}. Unjustified negligence by the
nonselfadjoitness of the Bogoliubov equations often lead to the association between
frequencies and energies in a way standard for quantum mechanics based on selfadjoint
operators, thus admitting possibility of complex energies in a conservative system.

However, rigorously proved theorem allowing clear physical interpretation of complex
frequencies is known from  the general theory of Hamiltonian systems  \cite{mackay}. It states
that excitations with complex frequencies   can appear only as a
result  of the resonance between two  excitations with positive and negative
energies \cite{sign}. The roots of this theorem go back to 19th century \cite{weir} 
and it has been used for interpretation of instabilities 
with complex eigenvalues in plasma physics
\cite{hasegawa}  and  fluid dynamics \cite{saffman}.
Motivated by recent experiments on observation of vortices in binary mixture of condensates
\cite{matthews} we will demonstrate that these vortices can have complex frequencies in their
spectrum, thereby  giving  good practical ground for  
selfconsistent theoretical interpretation of this phenomenon  in the BEC context.

Properties of  vortices in  the single component magnetically trapped ultracold  gases have been subject to the intensive  theoretical investigations, see e.g. [9,17-32], which significantly extended classical works 
\cite{pit,fetter} dealing with spatially unbounded case.  The properties of unit vortices in the two component condensates have also been studied, in parallel and independently from this work,  
by Garcia-Ripoll and P\'erez-Garcia \cite{ripoll1,ripoll2}. 
Richness of the dynamics of the two-condensate system and different approaches to the problem have led only to few overlaps  which are outlined where appropriate.

In the next section we introduce coupled GP equations and briefly describe their general properties.  Then, in section III, we derive Bogoliubov equations for excitations, clearly specifying differences between frequency and energy spectra of the excitations. We also explain  scenario of appearance  of the excitations with complex frequencies and show that they have zero energies. In section IV we verify validity of general results presented in Sec. III using perturbation theory in the limit of weak interaction and direct numerical study of Bogoliubov equations.  In Sections IV, V we describe how long term dynamics of unit and higher order vortices can be interpreted using linear Bogoliubov theory.

%****************************************************
\section{Gross-Pitaevskii equations}
%**************************************************
Studies of superfluid mixtures  using coupled GP equations  have long history and attracted significant recent activities, see e.g.
\cite{matthews,goldstein,phasesep,ripoll1,ripoll2,colson,shenoy,law,esry,ohberg} 
and references therein. Following
these works we assume that wave functions $\psi_{1,2}$ of two-species condensate inside an
axial harmonic trap  obey  equations 
%CHECK THESE EQUATIONS IN THE PROOFS
\begin{eqnarray}
\nonumber&&i\hbar\partial_t\psi_1=-\frac{\hbar^2}{2m}\vec\nabla^2\psi_1+
\frac{1}{2}m\Omega^2(r^2+\sigma^2z^2)\psi_1\\
&&+(u_{11}|\psi_1|^2+u_{12}|\psi_2|^2)\psi_1,\label{eq1}\\
\nonumber && i\hbar\partial_t\psi_2=-\frac{\hbar^2}{2m}\vec\nabla^2\psi_2+
\frac{1}{2}m\Omega^2(r^2+\sigma^2 z^2)\psi_2\\
\nonumber &&+
(u_{22}|\psi_2|^2+u_{21}|\psi_1|^2)\psi_2,
\end{eqnarray}
where for simplicity we have neglected by possible differences of
atomic masses $m_{1,2}=m$ and trap frequencies, 
$\Omega_{1,2}=\Omega$, $\sigma_{1,2}=\sigma$, 
$\vec\nabla=\vec i_x\partial_x+\vec i_y\partial_y+\vec i_z\partial_z$, $r^2=x^2+y^2$.
Coefficients $u_{ij}=4\pi\hbar^2a_{ij}/m$ characterise intra- and inter-species
interaction with corresponding two-body scattering lengths $a_{11}\ne a_{22}$
and $a_{12}=a_{21}$. 

At this point we introduce dimensionless time and space variables
$\tilde t=\Omega t$ and $(\tilde x,\tilde y,\tilde z)=(x,y,z)/a_{ho}$
and normalisation for the wave functions
$\tilde\psi_{1,2}=a_{ho}^{3/2}\psi_{1,2}/\sqrt{N_{1}}$,
where $a_{ho}={\sqrt{\hbar/(m\Omega)}}$ is the harmonic oscillator strength and 
$N_{1,2}=\int dV|\psi_{1,2}|^2$  are the numbers of particles.
We will consider the quasi-2D model to simplify our numerical study. 
This approximation was previously used in several works, 
see e.g. \cite{pu99,ripoll1,ripoll2}, and it is applicable not only  
for pancake traps, $\sigma\gg 1$, but also captures main qualitative 
features of spherical traps. 
To make further reduction of Eqs. (\ref{eq1}) we redefine  the  wave functions once more:
\begin{equation}
\tilde\psi_{1,2}=\left[\frac{\sigma}{2\pi}\right]^{1/4}\Psi_{1,2}
(\tilde x,\tilde y,\tilde t)
e^{-\sigma \tilde z^2/4}e^{-i\mu_{1,2}\tilde t-i\sigma \tilde t/2},\label{eq2}
\end{equation}
where $\mu_{1,2}$ are the chemical potentials. Dropping 
tilde we find that equations for $\Psi_{1,2}$ and $\Psi_{1,2}^*$  
can be put into Hamiltonian form
\begin{equation}
i\partial_t\vec\psi+\hat\eta{\delta H\over\delta\vec\psi^*}=0,
\label{ham1} 
\end{equation} 
$$\vec\psi=
\left[\begin{array}{c}
\Psi_1\\
\Psi_1^* \\
\Psi_2\\
\Psi_2^*
\end{array}\right],
\hat\eta=\left[\begin{array}{cccc}
-1&0&0&0\\
0&1&0&0 \\
0&0&-1&0\\
0&0&0&1
\end{array}\right],{\delta H\over\delta\vec\psi^*}=
\left[\begin{array}{c}
\delta H/\delta\Psi_1^*\\
\delta H/\delta\Psi_1 \\
\delta H/\delta\Psi_2^*\\
\delta H/\delta\Psi_2
\end{array}\right],$$
\begin{eqnarray}
H&&=\int dxdy(|\vec\nabla\Psi_1|^2+|\vec\nabla\Psi_2|^2\label{ham2}\\
\nonumber &&+(\hat V- \mu_1)|\Psi_1|^2 +(\hat V-\mu_2)|\Psi_2|^2\\
\nonumber && +{g\over 2}[\beta_{11}|\Psi_1|^4+\beta_{22}|\Psi_2|^4+
2\beta_{12}|\Psi_1|^2|\Psi_2|^2]).
\label{ha}\end{eqnarray}
where $H$ is the energy functional (or Hamiltonian), 
$\vec\nabla=\vec i_x\partial_x+\vec i_y\partial_y$, $\hat V=r^2/4$ is
the  harmonic potential, $g$ is the interaction parameter 
$g=8{\sqrt{\pi\sigma}}N_1a_{11}/a_{ho}$,  
$\beta_{12}=a_{12}/a_{11}$, and $\beta_{22}=a_{22}/a_{11}$.
$\beta_{11}=1$ and it is left in the equations for the sake of the symmetry. 

Invariancies of $H$ with respect to the infinitesimal 
 rotations  and two parameter phase transformation
\begin{equation}
(\Psi_1,\Psi_2)\to(\Psi_1e^{i\phi_1}, \Psi_2e^{i\phi_2}), \label{sym1}
\end{equation}
result in conservation of the total angular momentum and of 
the total number of particles in each component. 

Radially symmetric stationary states of the condensate (equilibria) 
can be presented in the form
\begin{equation}
\Psi_{j}=A_j(r)e^{iL_j\theta}, ~j=1,2
\end{equation} where $\theta$ is the polar angle and $A_j$ are real functions.
 Using  method suggested in \cite{matthews,Williams} 
only  states with vortex in one of the components can be  created 
and therefore   we will consider below only cases with $L_2> 0$ and $L_1=0$. 
Functions $A_j(r)$  were found numerically using Newton method.
Chemical potentials $\mu_{1,2}$ were found from  
the normalization conditions 
\begin{equation}
2\pi\int rdr A_{1}^2=1, 
~2\pi\int rdr A_{2}^2=\frac{N_2}{N_1}\equiv N.\label{norm0}
\end{equation}

%******************************************************************
\section{Frequency and energy spectra of collective excitations}
%*****************************************************************
\subsection{Bogoliubov equations and frequency spectrum}
%*****************************************************************
To study spectrum of BEC at equilibrium 
we linearise  Eqs. (\ref{ham1})  using  substitutions
\begin{equation}
\Psi_j=(A_j(r)+f_j(r,\theta,t))e^{iL_j\theta}, \label{eq5}
\end{equation}
where $f_j$ are small and  complex. 
Assuming that excitation are periodic in $\theta$ with period $2\pi$ we expand
$f_j$ into Fourier series: 
\begin{equation}
f_j=\sum_l\left(U_{lj}(r,t)e^{il\theta}+ V^*_{lj}(r,t) e^{-il\theta}\right),
\label{fexpan}\end{equation}
where $l=0,\pm 1,\pm 2,\dots$ Then
\begin{equation}
i\partial_t\vec W_l+\hat\eta\hat{\cal H}_l \vec W_l=0\label{eq6}
\end{equation}
is the set of linear partial differential equations resulting from the
substitution of (\ref{eq5}), (\ref{fexpan}) into Eq. (\ref{ham1}).
Here $\vec W_l=(U_{l1},V_{l1},U_{l2},V_{l2})^T$, 
$$\hat{\cal H}_l= \left[\begin{array}{cccc} 
\hat {\cal L}_{l,1} &   g\beta_{11}A_1^2 &g\beta_{12}A_1A_2&g\beta_{12}A_1A_2  \\
 g\beta_{11}A_1^2 &\hat {\cal L}_{-l,1} & g\beta_{12}A_1A_2&g\beta_{12}A_1A_2  \\
g\beta_{12}A_1A_2 &g\beta_{12}A_1A_2 &\hat{\cal  L}_{l,2} & g\beta_{22}A^2_2  \\
 g\beta_{12}A_1A_2 &g\beta_{12}A_1A_2 &g\beta_{22}A_2^2&\hat {\cal L}_{-l,2} 
 \end{array}\right],$$   
is a self-adjoint operator and 
\begin{eqnarray}
\nonumber&&\hat{\cal L}_{l,j}
=-\frac{1}{r}\frac{\partial}{\partial r}r\frac{\partial}{\partial r}
+\frac{1}{r^2}(L_j+l)^2+\hat V
-\mu_j \\
&&\nonumber +2g\beta_{jj}A_j^2+g\beta_{12}A_{j^{\prime}}^2;~j=1,2;~j^{\prime}=2,1.
\end{eqnarray}
Frequency spectra of  $\hat\eta\hat{\cal H}_l$ are discrete, providing $\hat V\ne
0$, therefore phonons, strictly speaking, are absent in the trap geometry.   In
accord with standard terminology \cite{pit,book}, all spatially bounded elementary
excitations can be called  collective excitations (or collective modes) of an
equilibrium under consideration. Linearised equations for  excitations in a Bose
gas, similar to Eq. (\ref{eq6}),  were first derived by Bogoliubov and  expansion
(\ref{fexpan}) was first applied in  the context of the vortex excitations by
Pitaevskii  \cite{pit}. To find frequency spectrum and collective modes we need to
solve set of the eigenvalue problems 
\begin{equation} \hat\eta\hat{\cal H}_l\vec w_{ln}=\omega_{ln}\vec w_{ln}.\label{evp} 
\end{equation}
$\hat\eta\hat{\cal H}_l$ are non-self-adjoint operators and therefore  complex
frequencies are not forbidden. If $\vec w_{ln}$ is an eigenvector of $\hat
\eta\hat{\cal H}_l$ with eigenvalue $\omega_{ln}$ it is selfevident that $\vec
w^*_{ln}$ is also an eigenvector with eigenvalue $\omega^*_{ln}$ and it can be
shown  that $\hat\eta\hat{\cal H}_{-l}$   has eigenvectors $\vec w_{-ln}=\hat\tau\vec
w_{ln}$ and $\vec w_{-ln}^*$  with eigenvalues $-\omega_{ln}$ and
$-\omega^*_{ln}$, respectively. Here 
$$\hat\tau=
\left[\begin{array}{cccc}
0&1&0&0\\
1&0&0&0\\
0&0&0&1\\
0&0&1&0
\end{array}\right].
$$
Thus spectrum of $\hat\eta\hat{\cal H}_{-l}$  can be obtained by reflection of the
spectrum of $\hat\eta\hat{\cal H}_{l}$  with respect to the line $Re\omega=0$ in
the  plane $(Re\omega,Im\omega)$. In other words, it means that purely real or
purely imaginary frequencies of the elementary excitation  exist in pairs and
complex frequencies  exist in quartets  and that
\begin{equation}
Tr\{\hat\eta(\hat{\cal H}_l+\hat{\cal H}_{-l})\}=\sum_n(\omega_{ln}+\omega_{-ln})=0. \label{tr}
\end{equation} 
Any equilibrium state of the condensate is {\em spectrally} stable \cite{mackay}
if its spectrum  is real. If there is at least one frequency with negative
imaginary part then corresponding collective mode will grow in time destabilising the equilibrium, which is called spectrally unstable.

%****************************************
\subsection{Biorthogonality}
%****************************************
Eigenmodes of $\hat\eta\hat{\cal H}_l$ are biorthogonal to  the modes of the 
adjoint eigenvalue problem \cite{biorth}, i.e.
\begin{equation}
\langle\vec w_{ln},\vec a_{ln^{\prime}}\rangle=0
\label{norm2},\end{equation}
where $n\ne n^{\prime}$ and $\vec a_{ln}$ obey
\begin{equation}
(\hat\eta\hat{\cal H}_l)^{\dagger}\vec a_{ln}
=\hat{\cal H}_l\hat\eta\vec a_{ln}
=\omega^*_{ln}\vec a_{ln},
\label{ajoint}\end{equation} 
 and $\langle \vec b,\vec
c\rangle=2\pi\sum_k\int_0^{\infty}rdr  b^*_{k} c_{k}$  for any  $\vec b$, $\vec
c$. Factor $2\pi$ is introduced to mimic integration over $\theta$. 
The key feature of our model, originating in its Hamiltonian structure,  
is that transformation linking $\vec w_{ln}$ and its
adjoint $\vec a_{ln}$ can be found in explicit and simple form.
If $\vec
w_{ln}$ is an eigenmode of  $\hat\eta\hat{\cal H}_l$ with frequency $\omega_{ln}$
then it can be checked that  $\hat\eta\vec w_{ln}^*$ and $\hat\eta\vec w_{ln}$ are
eigenmodes of $\hat{\cal H}_l\hat\eta$ with eigenvalues, respectively,
$\omega^*_{ln}$ and $\omega_{ln}$. The mode adjoint to $\vec w_{ln}$ is $\hat
\eta\vec w_{ln}^*$, therefore if $Im\omega_{ln}\ne 0$ then biorthogonality condition (\ref{norm2}) implies
\begin{equation}
\langle\vec w_{ln},\hat\eta\vec w_{ln}\rangle=0.
\label{norm3}\end{equation}
If $\omega_{ln}$ is real, then $\vec w_{ln}$ is also  real and 
$\langle\vec w_{ln},\hat\eta\vec w_{ln}\rangle\ne 0$.
Normalisation constant can always be chosen in such a way that
\begin{equation}
|\langle\vec w_{ln},\hat\eta\vec w_{ln}\rangle|=2.\label{norm4}
\end{equation}
Convenience of making the left hand side of Eq. (\ref{norm4}) equal to $2$ will become clear below,  when Eqs. (19), (20) for energies of elementary excitations are derived.
Eq. (\ref{norm4}) makes it explicit that inner product $\langle\vec w_{ln},\hat\eta\vec w_{ln}\rangle$ can be either positive or negative.
This point often remains silent if one derives conditions similar to (\ref{norm4})  
as part of the diagonalisation procedure of the second-quantised Hamiltonian disregarding eigenmodes with negative and zero values of $\langle\vec w_{ln},\hat\eta\vec w_{ln}\rangle$  \cite{fetter}. 
The fact that inner product 
of a real eigenmode with its adjoint can be negative,
is different from standard quantum mechanics based on the  selfadjoint operators.  These
difference can have series of consequences and one of them is that energy levels
are not necessarily linked to the eigenfrequencies according to the standard rule $\hbar\omega$, see below.
Note, that  origin of the nonselfadjoitness in our case is the nonlinearity
of  GP equations. If particles in the condensate do not interact, $g=0$, then
$\hat\eta\hat {\cal H}_l$ become diagonal and  self-adjoint.

%**************************************
\subsection{Energy spectrum}   
%**************************************
As a prelude to calculation of energies of the elementary excitations it is
instructive to introduce notion of the {\em nonlinear stability}, i.e. stability under the full nonlinear dynamics. According to
Dirichlet's theorem \cite{mackay}, nonlinear stability  is ensured 
if  an equilibrium  state under consideration is either minimum or maximum of
the functional $H$. Note here,  that excitations change number 
of particles in the equilibrium state, therefore   it was convenient 
to introduce energy functional $H$, which is actually the so-called 
modified energy \cite{fetter}, i.e.  it is the energy functional for 
Eq. (1) modified by addition of the number of particles integrals, 
see terms proportional to $\mu_j$ in Eq. (\ref{ha}). 
Let us assume that $\vec w_l(r)e^{il\theta}$ is a small 
initial perturbation of an equilibrium  state of the condensate, then 
\begin{equation}
H=H_0+{1\over 2}\langle\vec w_l,\hat{\cal H}_l\vec w_l\rangle+\dots,
\label{varh}\end{equation}
where $H$ is the energy of the perturbed equilibrium and $H_0$ is  calculated at
the exact equilibrium. The equilibrium is nonlinearly stable  if eigenvalue problems
\begin{equation}
 \hat{\cal H}_l\vec\beta_{lm}=\alpha_{lm}\vec\beta_{lm}
 \label{sa}\end{equation}
have all their eigenvalues either negative or positive, except  zero eigenvalues generated by continuous symmetries. 
Spectral instability  implies nonlinear one, and nonlinear stability 
implies spectral one,  but not vise versa \cite{mackay}.  In the nonrotating traps all the higher
order states of the condensate  including vortices are not the local 
extrema of the energy \cite{rokshar,svidzinsky98,linn99,ripoll99}.
Let us stress, however, that their nonlinear instability can  not be 
guaranteed by this fact alone and requires separate consideration.

\begin{figure} \setlength{\epsfxsize}{14.0cm}
\centerline{\epsfbox{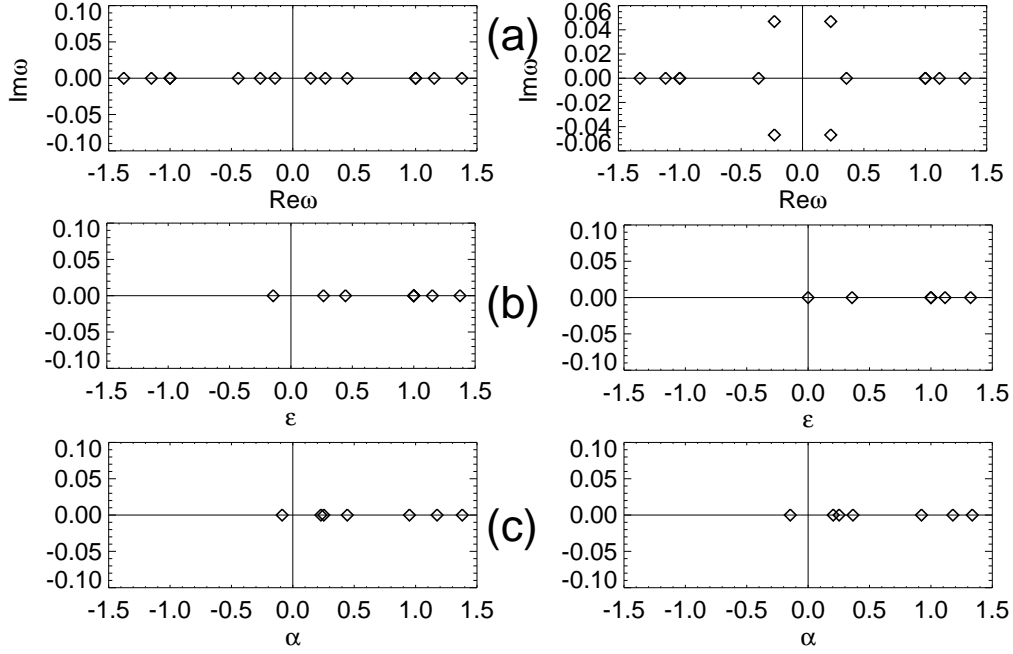}} 
\caption{Numerically calculated frequency (a) and energy (b) spectra 
corresponding to the collective modes with $l=\pm 1$ 
of the spectrally stable (left panel)
and spectraly unstable (right panel) unit vortices: $L_1=0$, $L_2=1$, $g=250$, $N=1$, 
other parameters for the left/right panel correspond to the vortex in the state 1/state 2,
see text after Eq. (32). 
The correspondence between frequency and energy spectra is obvious, see  Eqs.
(\ref{energies},\ref{energies2}).
(c) Spectrum of the eigenvalue problems 
$\hat{\cal H}_{\pm 1}\vec\beta_n=\alpha_n\vec\beta_n$; parameters as for (a),(b).}
\label{fig1} \end{figure}

If $\vec w_l$ in Eq. (\ref{varh}) is  an eigenmode $\vec w_{ln}$ 
of $\hat\eta\hat{\cal H}_l$ then 
\begin{equation}
\epsilon_{ln}={1\over 2}\langle\vec w_{ln},\hat{\cal H}_l\vec w_{ln}\rangle
={1\over 2}\omega_{ln}\langle\vec w_{ln},\hat{\eta}\vec w_{ln}\rangle
\label{eps}
\end{equation}
measures energy of this mode. Assuming that $\omega_{ln}$ is real 
and using biorthogonality conditions (\ref{norm4}) one gets
\begin{equation}
\epsilon_{ln}=\omega_{ln}sign\langle\vec w_{ln},\hat\eta\vec w_{ln}\rangle.
\label{energies}
\end{equation}
If $Im\omega_{ln}\ne 0$ then Eq. (\ref{norm3}) implies
\begin{equation}
\epsilon_{ln}=0. 
\label{energies2}
\end{equation}
Energy in physical units is given by $\epsilon_{ln}\hbar\Omega$.

It is  readily demonstrated that 
\begin{equation}
\langle\vec w_{ln},\hat\eta\vec w_{ln}\rangle 
=-\langle\hat\tau\vec w_{ln},\hat\eta\hat\tau\vec w_{ln}\rangle
=-\langle\vec w_{-ln},\hat\eta\vec w_{-ln}\rangle,
\end{equation}
therefore modes with frequencies $\omega_{ln}$ and $\omega_{-ln}=-\omega_{ln}$ 
have their energies equal in sign and value. 

Eigenfunctions $\vec\beta_{ml}$ are orthogonal
$\langle\vec\beta_{lm},\vec\beta_{lm^{\prime}}\rangle=\delta_{mm^{\prime}}$
 and form complete basis. Therefore $\vec w_{ln}$ in Eq. (\ref{eps}) 
can be expanded
in terms of $\vec\beta_{lm}$, which gives $\epsilon_{ln}=\sum_m\alpha_{lm}
\langle\vec w_{ln},\vec\beta_{lm}\rangle^2$.
Thus an equilibrium can have  collective modes carrying energy with opposite 
signs only if it is not a local extremum of the energy functional.

%**************************************************************
\subsection{Resonances and instabilities, crossings and avoided crossings}
%***************************************************************
Instabilities with imaginary eigenvalues have been found to dominate dynamics of 
homogeneous BEC with attractive interaction \cite{bog} and mixture of two condensates with repulsive interaction \cite{goldstein,phasesep}.  Higher order vortices and dark solitons in single component trapped condensates can be unstable with respect to modes with complex
frequencies \cite{pu99,federnew}. In our problem instabilities with complex eigenvalues  are the most important ones also. Therefore it is desirable to have criterion or theorem allowing 
interpretation and prediction of these instabilities. Such theorem is actually known from the general theory of the Hamiltonian dynamical systems \cite{mackay} and, in our context, it can be reformulated as:   {\em (i)~Sign of the energy of a collective excitation preserves as parameters vary as long as there is no frequency resonance with
another excitation.  (ii)~Condensate at equilibrium can lose spectral stability as parameters vary only in two ways: either  via frequency resonance of  two  elementary excitations with positive and negative energies or by resonance  at zero frequency.} Proof
of these results \cite{mackay} is based on the fact, that transition from spectral stability to
instability can not violate energy conservation law: $\partial_tH=0$.

From this theorem and preceding considerations one can conclude  that complex eigenvalues
in the spectrum of the vortices in Bose-Einstein condensate can appear only due to mutual annihilation of the collective excitations with positive and negative energies.  The latter can coexist only for nonlinearly unstable equilibria. The theorem does not forbid
for frequencies of two excitations with either the same or opposite signs of the energies simply cross each other without change of the spectral stability. In our example it typically happens when corresponding eigenmodes remain orthogonal at the exact resonance.
If, however, eigenmodes of the colliding excitations start to compete for the same direction in the functional phase space and become degenerate at the resonance then crossing is not a generic scenario. If the signs of the energies of the excitations are
opposite, then quartet of complex frequencies appears upon passing the resonance. Alternatively, if the signs are the same,  the exact resonance can not be achieved and it
becomes replaced by the so-called  avoided crossing of the  energy levels \cite{mackay}. 

Figs. 1(a),(b) show numerically calculated frequency and energy spectra 
of the collective excitations with
$l=\pm 1$ for spectrally stable (left panel) and spectrally unstable (right panel) unit
vortices $(L_1=0,~L_1=1)$. 
The negative energy excitation is clearly seen in the spectrally stable situation. At the transition
threshold to spectral instability this excitation  and another
one, having energy with the same absolute value but opposite sign, annihilate each other.
This transition is accompanied by appearance of the zero energy excitations.
 Examples of frequency and energy
evolution under the parameter variation resulting in instabilities, crossings and
avoided crossing are given in the  section IV.
%
%CHECK THIS PLACE IN THE PROOFS
Figs. 1(c)  show
spectra of $\hat {\cal H}_{\pm 1}$ illustrating that  there are no qualitative 
changes in these spectra after appearance of complex frequencies.

Possibility of observation of the collective modes with complex frequencies  has 
caused some  concerns and discussions \cite{federnew,ripoll99}. However, initial
perturbations of the  equilibrium  state having nonzero projections on the 
 adjoint mode $\hat\eta\vec w_{ln}^*$
with $Im\omega^*_{ln}>0$ will lead to the ultimate growth of the corresponding
mode $\vec w_{ln}$ because it has $Im\omega_{ln}<0$.
The consequences of this growth is in any way diminished by the fact that 
the energy of this mode is zero.
Wealth of references and examples of instabilities with complex eigenvalues existing in other physical contexts  can be found in \cite{mackay,skr}. 

%****************************************************
\subsection{Goldstone and dipole modes}
%*************************************************
Infinitesimal variations of $\phi_j$, see Eq. (\ref{sym1}), generate two  zero-energy eigenmodes
(Goldstone modes) $(A_1,-A_1,0,0)^T$ and $(0,0,A_2,-A_2)^T$ belonging to the null-eigenspace
of  $\hat\eta\hat{\cal H}_0$. Harmonic trapping  modifies spectrum  in such a way that  $\hat\eta\hat{\cal
H}_{\pm 1}$  acquire couple of parameter independent  eigenvalues $\omega=\pm 1$  with
eigenfunctions
 \begin{equation} \vec w_{\pm 1d}= \left[\begin{array}{c}
\cr\omega\left(\frac{d A_1}{d r}\mp \frac{1}{r}L_1A_1\right)  +\frac{1}{2}rA_1 \\
\cr\omega\left(\frac{d A_1}{d r}\pm \frac{1}{r}L_1A_1\right)-\frac{1}{2}rA_1 \\
\cr\omega\left(\frac{d A_2}{d r}\mp \frac{1}{r}L_2A_2\right) + \frac{1}{2}rA_2\\
\cr\omega\left(\frac{d A_2}{d r}\pm \frac{1}{r}L_2A_2\right)-\frac{1}{2}rA_2
\end{array}\right],\label{eq9}\end{equation} where $\omega$ can take values  $\pm 1$ for
both eigenmodes. 
$\vec w_{\pm1d}$  are often called dipole modes and  $\epsilon_{\pm 1d}=1$.  Eqs. (\ref{eq9})
generalise expressions  previously derived for single-species condensates
\cite{svidzinsky98}. Existence of the dipole modes can also be associated with the Kohn's
theorem \cite{kohn,dobson}. 

%*****************************************************
\section{Collective excitations of unit vortices: $L_1=0,~ L_1=1$}
%*****************************************************
We begin our analysis considering  weakly interacting condensate: $g\ll 1$, $N\sim 1$.  In
this limit potential energy due to harmonic trapping $\hat V$ strongly dominates over the 
interaction energy, which    allows to make explicit calculations of the frequency and
energy spectra of the elementary excitations. Calculations of the excitation
spectra of the vortices and dark solitons in the singly component weakly interacting
condensates have been recently done by several groups of authors \cite{federnew,linn99}.
However, these calculations lack  analysis of the energy sign in a sense explained in the
preceding section.  

We substitute asymptotic expansions
$\mu_{1,2}=\mu_{1,2}^{(0)}+g\mu_{1,2}^{(1)}+O(g^2)$, 
$A_{1,2}=A_{1,2}^{(0)}+gA_{1,2}^{(1)}+O(g^2)$ into the stationary ($\partial_t=0$) 
version  of Eqs. (\ref{ham1}) and derive recurrent system of linear equations. 
In the zero approximation we have two uncoupled harmonic oscillator problems
with eigenmodes \begin{equation}  
A_{1}^{(0)}=\frac{1}{{\sqrt{2\pi}}}e^{-r^2/4}, ~
A_{2}^{(0)}=\sqrt{\frac{N}{\pi}}\frac{r}{2}e^{-r^2/4}\label{eq12}.
\end{equation} 
Using solvability condition of the first order problem we find 
asymptotic expressions for the  chemical
potentials: $\mu_1=1+g(2\beta_{11}+n\beta_{12})/(8\pi)+O(g^2)$,
$\mu_2=2+g(n\beta_{22}+\beta_{12})/(8\pi)+O(g^2)$. 

Then we expand operators $\hat {\cal H}_l$,  eigenmodes $\vec w_l$ and frequencies
$\omega_l$ into the series $\hat {\cal H}_l=\hat {\cal H}_l^{(0)}+g\hat {\cal
H}_l^{(1)}+O(g^2)$, $\vec w_l=\vec w_l^{(0)}+g\vec w_l^{(1)}+O(g^2)$, 
$\omega_l=\omega_l^{(0)}+g\omega_l^{(1)}+O(g^2)$. After substitution into Eq. (\ref{evp})
in the first order we find standard equation
\begin{equation}
(\hat\eta\hat {\cal H}_l^{(0)}-\omega_l^{(0)})\vec w_l^{(1)}=
(\omega_l^{(1)}-\hat\eta\hat {\cal H}_l^{(1)})\vec w_l^{(0)}.
\label{perteq}\end{equation}
$\hat\eta\hat {\cal H}_l^{(0)}$ is diagonal and self-adjoint and all its eigenmodes and
eigenvalues can be found explicitly. Then using solvability condition for Eq.
(\ref{perteq}) one can find  corrections for all frequencies and coefficients in the linear
superposition of the zero approximation eigenmodes. Computer algebra makes  technical
realization of this plan by a straightforward exercise. We will present and analyse
explicit analytical  results only for the operator $\hat\eta\hat{\cal H}_{-1}$ considering
vicinity of the spectral point $\omega=1$, because it contains information about origin of
spectral instabilities of unit vortices. Equivalent analysis of $\hat\eta\hat{\cal H}_1$
near  $\omega=-1$ gives the same results.

$\hat\eta\hat{\cal H}^{(0)}_{-1}$ has  three unit frequencies, $\omega^{(0)}_{-1}=1$.
Corresponding eigenmodes are 
\begin{eqnarray}
\nonumber 
&& \vec b_{1}=(0,1,0,0)^T\frac{r}{{2\sqrt{\pi}}}e^{-r^2/4},\\
&&   \vec b_{2}=(0,0,1,0)^T\frac{1}{{\sqrt{2\pi}}}e^{-r^2/4},\label{vect}\\ 
\nonumber && \vec b_{3}=(0,0,0,1)^T\frac{r^2}{{4\sqrt{\pi}}}e^{-r^2/4}.
\end{eqnarray}
Solvability conditions for Eq. (\ref{perteq}) lead to  the characteristic determinant of
the three by three matrix. We find that one of the three frequencies is  associated with 
dipole mode $\vec w_{-1d}$ and that the other two  are 
\begin{eqnarray}
&&\omega_{-1}^{\pm}=1+\frac{g}{32\pi}\Big(-3\beta_{12}-N\beta_{22}
\pm \sqrt{R} \Big)+O(g^2),\label{ev1}\\
\nonumber &&R\equiv(3\beta_{12}+N\beta_{22})^2-8\beta_{12}^2(N+1).
\end{eqnarray}
Corresponding unnormalised eigenmodes are
\begin{eqnarray}
\nonumber \vec w_{-1}^{\pm}=&& \sqrt{2N}\left(N\beta_{22}-\beta_{12}
\pm \sqrt{R}\right)\vec b_{1}\\ 
\nonumber &&+ \sqrt{2}\left( 3N\beta_{22}+ [1-4N]\beta_{12}\mp \sqrt{R}\right)\vec b_{2}\\ 
&&+4N\left(\beta_{12}-\beta_{22}\right) \vec b_{3} +O(g).
\label{ev2}\end{eqnarray}

Pair of frequencies (\ref{ev1}) becomes  complex if $R<0$, signaling spectral instability
of the unit vortex. If $R>0$, then 
\begin{eqnarray}
&&\langle\vec w_{-1}^{\pm},\hat\eta\vec w_{-1}^{\pm}\rangle =\label{norm1}\\
\nonumber&& 4\left[R(N-1) \pm\sqrt{R}\left\{(1-5N)\beta_{12}+N(3+N)\beta_{22}\right\}\right] 
\end{eqnarray}
and one can check that \begin{equation}
\epsilon_{-1}^{\pm}=\pm\omega_{-1}^{\pm},\end{equation}
i.e.  modes with frequencies $\omega^{+}_{-1}$ and 
$\omega^{-}_{-1}$ have, respectively, positive and negative energies, see Fig. 2(c).
One can also verify biorthogonality  
$\langle\vec w^{\pm}_{-1},\hat\eta\vec w^{\pm}_{-1}\rangle=0=\epsilon^{\pm}_{-1}$  
for $R<0$, see Eq. (\ref{norm3}), and $\langle\vec w^{\pm}_{-1},\hat\eta\vec
w^{\mp}_{-1}\rangle=0$ for $R>0$, see Eq. (\ref{norm4}). 

Assuming $\beta_{ij}>0$ and rewriting instability condition $R<0$  in the form  
\begin{equation} \frac{\beta_{22}}{\beta_{12}}<\frac{1}{N}(2{\sqrt{2(N+1)}}-3),
\label{cond}\end{equation}
one can see that unit vortices are more
stable if intraspecies interaction of atoms in the vortex containing part of the 
condensate is somewhat larger than interspecies interaction.
The choices of scattering  lengths corresponding to the experiment \cite{matthews} are:
$\beta_{22}=1.03/0.97$, $\beta_{12}=1/0.97$ ($\beta_{22}>\beta_{11}$) and
$\beta_{22}=0.97/1.03$, $\beta_{12}=1/1.03$  ($\beta_{22}<\beta_{11}$).  The former case
corresponds to the vortex in the spin state  $\{ F=1,m_f=-1\}$ of $^{87}Rb$  and
the latter to the vortex in the state $\{ 2,2\}$.  These two states will be
called -- state  $1$ and state $2$. It is clear that for $N=1$ Eq. (\ref{cond}) predicts
instability for the vortex in the state $2$ and stability for the vortex in the state $1$,
which supports results of the   experimental observations \cite{matthews}.

Fig. 2 shows frequency resonance accompanied by the simultaneous
mutual annihilation of excitations with positive and negative energies happening at some critical value of  $\beta_{12}$, see Eq. (\ref{cond}).
In fact it models transition from the situation with vortex
in the state 1 to the case with vortex in the state 2. Performing
numerical studies for wide range of  parameters, outside the validity region of
analytical considerations, we have not been able to find regions  of spectral
instability of the unit vortex in  state $1$. Contrary, existence of  instabilities of the vortex in  state $2$  due to exactly the same scenario, which is predicted in the weak interaction limit, can be readily demonstrated, see  Fig. 3. 

It is important to stress,  that if two condensates are decoupled, $\beta_{12}=0$,  then the negative energy mode $\vec w_{-1}^{-}$ belongs to the vortex containing component  and the positive energy mode  $\vec w_{-1}^{+}$ belongs to the vortex free component. Because of this separation instability is not possible for any values of $g$, which agrees with spectral stability of unit vortices in the singly component case reported in \cite{pu99}. Thus we can conclude that the  vortex instability in our example has essentially two-component nature and its analog may also exist in the case when second component is a non-condensate one.

Criterion similar to (\ref{cond}), but without corresponding energy analysis, has also been independently obtained in  \cite{ripoll2} using two mode approach, i.e. condensate wave
functions $\Psi_1$ and $\Psi_2$ have been presented as linear superposition of modes (\ref{eq12}) with time dependent coefficients and GP equations have been reduced to the set
of ordinary differential equations for these coefficients \cite{ripoll2}. However, this
method fails taking into account an eigenmode proportional to $r^2$, see $\vec b_3$ in (\ref{vect}), which makes an important contribution to the expressions for frequencies.
Therefore Eqs. (\ref{ev1}) and (\ref{cond}) are  different from the corresponding results presented in \cite{ripoll2}. 

To illustrate crossings and avoided crossings in the spectrum of unit vortices we show in
Fig. 4 positive part of the frequency spectra of $\hat\eta\hat{\cal H}_{\pm 1}$ and $\hat
\eta\hat{\cal H}_{\pm 2}$. One can see numerous points, which on the first glance can be
interpreted as crossings in the frequency spectrum. However, under the close investigation
those of them which are marked by the open circles, turns to be the avoided crossing of the
excitations with equal  energy signs, see Fig. 5.

\begin{figure} \setlength{\epsfxsize}{7.0cm}
\centerline{\epsfbox{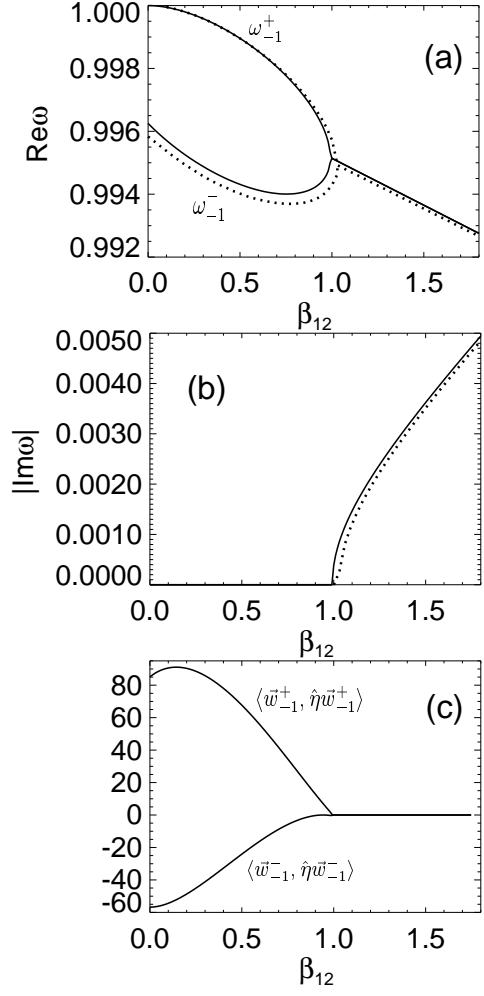}} 
\caption{(a),(b) Frequencies $\omega^{\pm}_{-1}$ vs $\beta_{12}$: $N=2$, $g=0.1$, $\beta_{22}=0.97/1.03$.
(c) Inner products $\langle\vec w_{-1}^{\pm},\hat\eta\vec w_{-1}^{\pm}\rangle$
characterising energy signs of the corresponding excitations, see Eq. (30).
Doted lines are
numerical solutions of $\hat\eta\hat{\cal H}_{-1}\vec w_l=\omega_l\vec w_l$ and full lines correspond to  Eqs. (28) and (30).} \label{fig2} \end{figure}
\begin{figure} \setlength{\epsfxsize}{7.0cm}
\centerline{\epsfbox{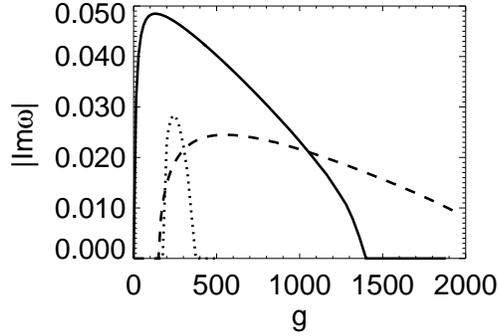}} 
\caption{Instability growth rate of the unit vortex in state 2 vs interaction parameter
$g$. Full line -- $N=1$; Dashed line -- $N=0.3$; Dotted line -- $N=9$.} 
\label{fig3} \end{figure}
\begin{figure} \setlength{\epsfxsize}{7.0cm}
\centerline{\epsfbox{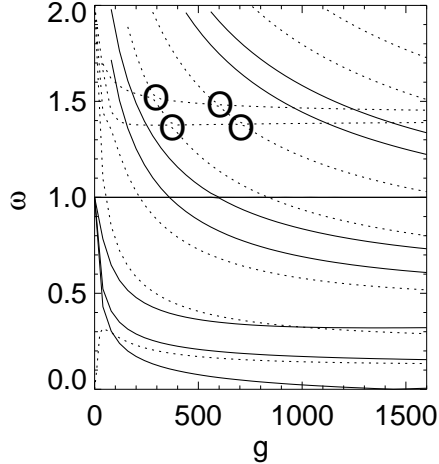}} 
\caption{Positive part of the frequency spectrum of the collective excitations 
with $|l|=1$ (full lines) and
$|l|=2$ (doted lines) of the unit vortex in state 1
vs interaction parameter $g$: $N=1$. Open circles mark the avoided crossings (i.e. avoided
frequency resonances) of the excitation with positive energies.} \label{fig4} \end{figure}

\begin{figure} \setlength{\epsfxsize}{7.0cm} 
\centerline{\epsfbox{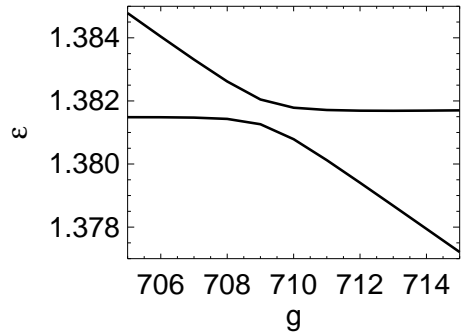}}
\caption{Dependencies of energies of two elementary excitations with $l=2$
vs $g$ showing details of the rightmost from the avoided crossings marked in Fig. 4.}
\label{fig5} \end{figure}

%*****************************************
\section{Drift of unit vortices}
%*****************************************
The unstable modes of the vortex in the state $2$  are of the dipole type, i.e. $|l|=1$, therefore their
growth leads to the initial displacement of the vortex from the trap center.  Displaced
vortex  carries  less of the total angular momentum, compare to the momentum of the vortex
positioned at the trap center. Lack of the angular momentum is compensated by two factors.
First, vortex acquires a nonzero tangential velocity and therefore its trajectory is
actually a spiral, which is similar to the dynamics of
an optical vortex  displaced from the center of a gaussian beam \cite{kivshar98oc}.
Second, angular momentum,  and vortex itself, become gradually
transferred into  the second condensate component, which was  first demonstrated
 in  \cite{ripoll1,ripoll2}.  

If drift instability is absent, then dissipative effects still can result in the vortex
drift, which was predicted for the single component condensates by several authors
\cite{rokshar,fedichev99}. Presence of the second condensate opens a channel for the 
energy and angular  momentum transfer from the vortex into the vortex free component.   It
can be seen that vector $\vec b_1$, being excited by the instability, provides a channel
for this transfer. Thus drift instability can also be interpreted as due to dissipation of
the energy and momentum by the vortex free condensate component.

In the limits $N\gg 1$ and $N\ll 1$ our model can be approximately  considered as a single component  condensate with ($N\gg 1$) or without ($N\ll 1$) vortex.  Unit vortex and ground state of the single component condensate are known to be spectrally stable. Therefore drift instability  disappears  in both limits, see Fig. 3. Increase of $g$ for fixed $N$ also results in the suppression of the instability, see Fig. 3,  which means that not only
relative, but also absolute increase of the number of atoms in the vortex free component
stabilizes the condensate.  

%************************************************************
\section{Drift and splitting of higher order vortices}
%***********************************************************
Considering higher order vortices one can expect to find  instability scenario resulting in
their splitting into unit vortices. This scenarion is expected to be due to growth of the
collective modes with $|l|>1$. However, as we will see below the drift instability linked
to the dipole like modes, $|l|=1$,  also can be presented. It leads to  displacement of the
whole vortex  from the trap center  without splitting, at least at the onset of the
instability. 

Both drift and splitting instabilities of the higher-order vortices appear as a result of
the frequency resonances of the elementary excitations with negative and positive energies,
similar to the case of  unit vortices. $N\ll 1$  corresponds  to the vortex free
condensate and therefore both instabilities disappear in this limit. In the limit
$N\gg 1$ only drift instability is suppressed and  one can  recover periodic in $N$  bands of the  instabilities with complex frequencies and $|l|>1$ similar to the results reported for
higher order vortices in a singly component condensate \cite{pu99}. Note, that  splitting
itself was not explicitly demonstrated in \cite{pu99}. It is also interesting to note that
higher-order vortices in the free, $\hat V=0$, single component condensate are spectrally
stable  \cite{aranson96}. Thus splitting  can be considered as induced by the trapping. The
vortex free condensate component plays crucial role in the drift instability,  but will the
latter one be presented without trapping  or not remains an open problem. 

\begin{figure}
\setlength{\epsfxsize}{7.0cm}
\centerline{\epsfbox{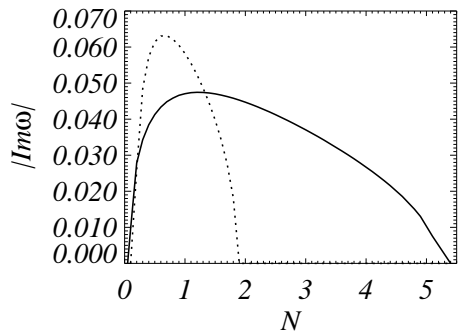}}
\caption{Growth rate of the drift $|l|=1$ (full lines) and splitting  $|l|=2$
(dotted lines) instabilities of the double vortex 
in the state $2$ vs $N$ for  $g=900$.}
\label{fig6}
\end{figure}

%*******************************************************
\subsection*{Double vortices: $L_1=0,~L_2=2$}
%*****************************************************
Considering double vortex we have found that it can be unstable with respect to the
$|l|=1,2$ excitations. The vortex in the state $1$ has been found  surprisingly stable. One
has to take relatively small values of $g$ and  large $N$ to find splitting instability.
Vortex in the state $2$ is more unstable in a sense that splitting exists already for
$N\sim 1$, see Fig. 6. As it is evident from Fig. 6 either drift or splitting instability
can dominate vortex dynamics. If drift instability is dominant,  then vortex first gets
displaced from the trap center and only then splits into the unit ones, see  Fig. 7. The
latter happens due to  the curved background which  breaks cylindrical symmetry with
respect to the vortex axis. After the splitting vortices remain close to each other and
move towards the condensate periphery. Results of the numerical 
simulation of GP equations (1) presented in this Section were
obtained  starting from equilibrium states perturbed by random noise
with amplitude $\sim 0.05A_{1,2}$.

The dynamics is quite different when splitting instability is dominant, see Fig. 8. In this
case unit vortices appear straight at the onset of the instability development and spiral
out  of the condensate center being always positioned symmetrically with respect to it.  
After a certain period of time vortices move back to the trap center and condensate state
close to the initial one is restored, see Fig. 8,  then the cycle is repeated with
gradually worsening degree of periodicity.

During development of the instability angular momentum and vorticity become partially
transferred into the second condensate. Analysis of the transverse profiles of the phases
corresponding to the density profiles shown in  Fig. 8, 
has revealed that black spots appearing in the second condensate are indeed unit vortices,
 not the  density holes without topological structure. 

\begin{figure}
\setlength{\epsfxsize}{18.0cm}
\centerline{\epsfbox{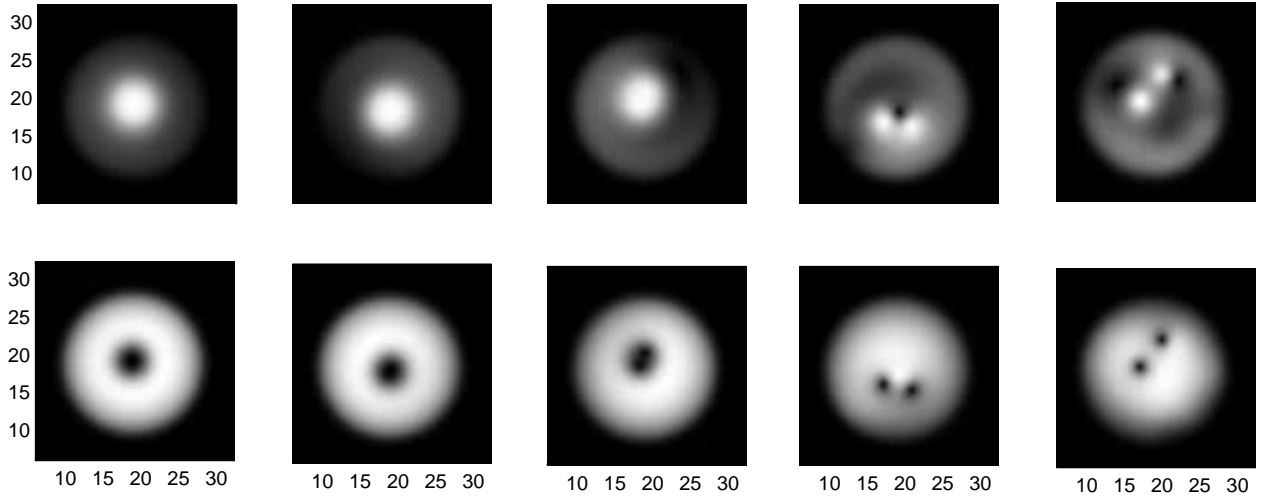}}
\caption{Development of the drift instability of the double vortex in the state $2$: 
$g=900$, $N=2.5$. Top  -- $|\Psi_1(x,y)|^2$, bottom  -- $|\Psi_2(x,y)|^2$.
Time interval between snapshots is $20$. First snapshot corresponds to 
$t=60$. }
\label{fig7}
\end{figure}

\begin{figure}
\setlength{\epsfxsize}{18.0cm}
\centerline{\epsfbox{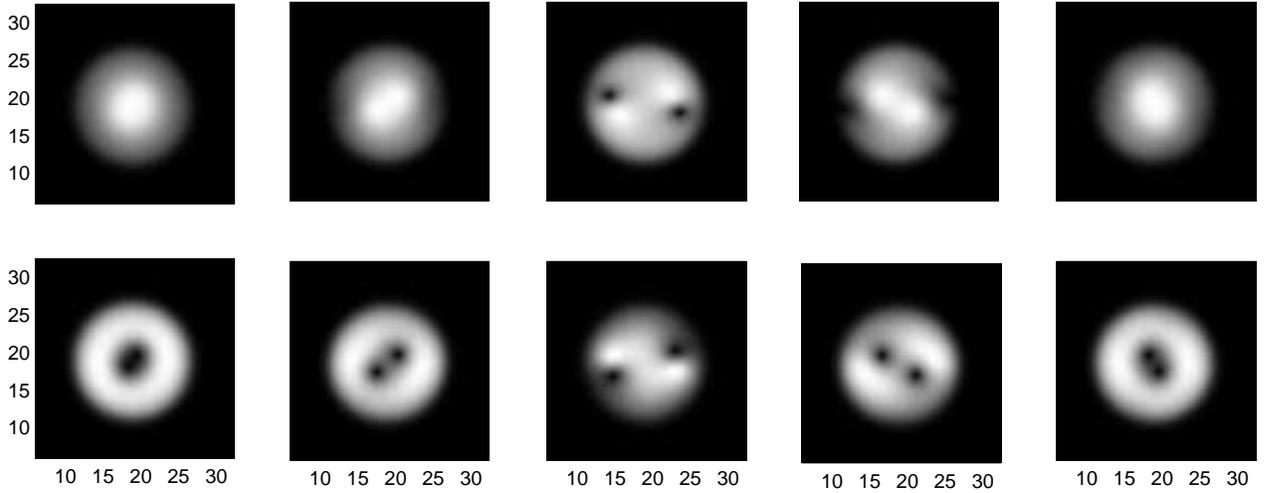}}
\caption{Development of splitting instability of the 
double vortex in the state $2$: 
$g=900$, $N=0.7$. Top  -- $|\Psi_1(x,y)|^2$, bottom  -- $|\Psi_2(x,y)|^2$.
Time interval between snapshots is $20$.
First snapshot corresponds to  $t=30$. }
\label{fig8}
\end{figure}

%*******************************************************
\subsection*{Triple vortices: $L_1=0,~L_2=3$}
%*******************************************************
Reach variety of beautiful vortex lattices can be found considering instabilities of
vortices of the order 3 and higher. This reachness can be understood in terms of spatial
profiles of the unstable collective modes. E.g. triple vortex has been found unstable with
respect to the perturbations with $|l|=1,2,3$. All components of the $|l|=2$ excitations
are equal to zero at the trap center. Therefore one can expect that growth of this mode will
develop into a spatial structure  preserving vortex at the trap center.  Contrary some of
the components of the $|l|=3$ modes have humps at the center. Therefore their growth should
repel all unit vortices out of the center, which leads to the  breaking of the triple
vortex into  a triangular structure of the unit vortices  moving away from the trap center.
Both instabilities can be found for the same parameters values and can have close growth rates. Therefore the winning mode  is selected through the process of complex competition, see Figs. 9,10.

\begin{figure}
\setlength{\epsfxsize}{7.0cm}
\centerline{\epsfbox{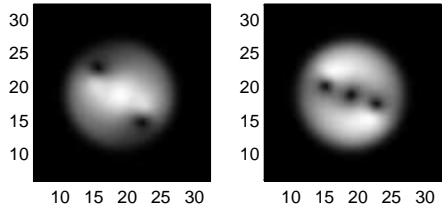}}
\caption{$|l|=2$ instability of the triple vortex in the state $2$: 
$g=900$, $N=0.9$, $t=70$. Left  -- $|\Psi_1(x,y)|^2$, right  -- $|\Psi_2(x,y)|^2$.
}
\label{fig9}
\end{figure}

\begin{figure}
\setlength{\epsfxsize}{7.0cm}
\centerline{\epsfbox{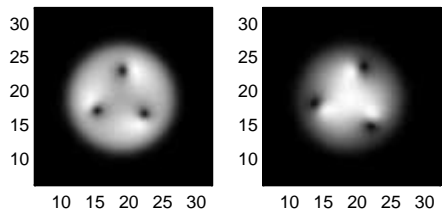}}
\caption{$|l|=3$  instability of the triple vortex in the state $2$: 
$g=900$, $N=0.9$, $t=85$. Left  -- $|\Psi_1(x,y)|^2$, right  -- $|\Psi_2(x,y)|^2$.
}
\label{fig10}
\end{figure}
%*****************************
\section{Summary}
%*****************************
We have described general approach to stability of equilibrium in  BEC using Bogoliubov theory and GP equation. Biorthogonality conditions (\ref{norm3}), (\ref{norm4}) and correspondence between frequency and energy spectra of the elementary excitations 
(\ref{energies}), (\ref{energies2})  have been derived selfconsistently from first principals revealing several novel and conceptually important aspects originating in nonseldjoitness of the Bogoliubov operator. 

It has been demonstrated that 
frequency resonances of the excitations  with positive  and negative energies can lead to their mutual annihilation  and appearance of the collective modes with complex frequencies and zero energies. Conditions for the avoided crossing of energy levels have also been discussed. General theory has been verified  
both  numerically and analytically in the  weak
interaction limit considering an example of vortices in a binary mixture of condensates.

Growth of excitations with complex frequencies  leads to the two main 
scenarios of the instability development. First one is the  spiraling of unit and double
vortices out of the condensate  center to its periphery. Second scenario is the  splitting 
of the double and higher order vortices into  unit ones.   
Absolute and/or relative increase of the
number of particles  in the vortex free condensate component have been found to have
stabilizing effect.

\acknowledgments
Author is particularly grateful to referees for their critical and very helpful comments.
He also acknowledges  discussions with S.M. Barnett and W.J. Firth. Numerical part of the
work was significantly speeded up due to access to the computer equipment obtained via U.K.
EPSRC grant GR/M31880 and assistance of G. Harkness and R. Martin.

%\newpage

%\end{multicols} 
%\newpage 

\end{document}